# High-Field Magnetization and Magnetic Phase Diagram of Metamagnetic Shape Memory Alloys $Ni_{50-x}Co_xMn_{31.5}Ga_{18.5}$ (x = 9 and 9.7)

T. Kihara[1]*, X. Xu[2], A. Miyake[3], Y. Kinoshita[3] M. Tokunaga[3], Y. Adachi[4], and T. Kanomata[5]

[1] Institute for Materials Research, Tohoku University, Sendai, Miyagi, Japan

[2] Department of Materials Science, Tohoku University, Sendai, Miyagi, Japan

[3] The Institute for Solid State Physics, The University of Tokyo, Kashiwa, Chiba, Japan

[4] Graduate School of Science and Engineering, Yamagata University, Yonezawa, Yamagata, Japan

[5] Research Institute for Engineering and Technology, Tohoku Gakuin University, Tagajo, Miyagi, Japan

* Address all correspondence to: t_kihara@imr.tohoku.ac.jp

**Abstract**

Magnetic phase diagrams of the metamagnetic shape memory alloys $Ni_{50-x}Co_xMn_{31.5}Ga_{18.5}$ (x = 9 and 9.7) were produced from high-field magnetization measurements up to 56 T. For both compounds, magnetic field induced martensitic transformations are observed at various temperatures below 300 K. Hysteresis of the field-induced transformation shows unconventional temperature dependence: it decreases with decreasing temperature after showing a peak. Magnetic susceptibility measurement, microscopy, and X-ray diffraction data suggest a model incorporating the magnetic anisotropy and Zeeman energy in two variants, which qualitatively explains the thermal and the magnetic field history dependence of the hysteresis in these alloys.

Triggered by the discoveries of magnetic-field-induced shape memory effect and inverse magnetocaloric effect (IMCE) in Ni-Mn-Z (Z = In, Sn, and Sb) Heusler alloys [1–3], investigators have studied Ni-Mn based metamagnetic shape memory alloys (MMSMAs) extensively during the last decade as a class of promising multi-functional materials offering potential for use as high-performance actuators and magnetic refrigerants functioning at around room temperature. These materials undergo martensitic transformation (MT) from ferromagnetic austenitic phase (A phase) to paramagnetic or weak-magnetic martensitic phase (M phase) concomitantly with decreasing temperature accompanied with the simultaneous change of electron density of states, crystal structure, and magnetic structure [4–11]. By applying a magnetic field to the M phase, the magnetic-field-induced martensitic transformation (MFIMT) is realized. Therefore, electronic, structural, and magnetic properties in the Ni-Mn based MMSMAs are controllable by the externally applied magnetic field, which is crucially important for development of functional materials. However, the large thermal hysteresis of MFIMT, which causes a large irreversible energy loss, is an obstacle for practical applications. Moreover, the underlying mechanism of thermal and magnetic field hysteresis of MT remains unclear.

Thermodynamically, the hysteresis related to the first-order phase transitions including MT results from the existence of an energy barrier between two phases. For MTs, this energy barrier is attributed mainly to the friction in the habit plane motion. Because MT is the diffusionless structural phase transition, the MFIMT can be regarded as a phenomenon that is analogous to plastic deformation [12–15]. According to Seeger and Kocks *et al.*, the critical shear stress of the plastic deformation can be divided into athermal and thermally activated terms [14,15]. Here, the athermal term is temperature-independent. Consequently, the temperature dependence of the critical shear stress is attributed to the thermally activated term, as confirmed experimentally by several works [16–19].

Temperature dependence of the critical shear stress was proposed theoretically based on the well-known Arrhenius relation by Ghosh *et al* [13]. Based on an analogy between plastic deformation and the MFIMT, Umetsu *et al*. proposed a model explaining the temperature dependence of the magnetic field hysteresis of $Ni_{50}Mn_{34.4}In_{15.6}$ [16]. In their model, the magnetic field hysteresis $H_{hys}$ is

$$H_{hys}(T) = H_\mu + H_{TA}(T)$$

$$= H_\mu + H_{TA}(0)\left[1 - \left(\frac{mk_BT}{\Delta}\right)^{1/q}\right]^{1/p}. \tag{1}$$

Here, $H_\mu$ and $H_{TA}$ respectively denote the athermal and the thermally activated contributions. $k_B$ and $\Delta$ respectively represent Boltzmann's constant and the average height of the rigid energy barrier between A and M phases at 0 K. Also, $m = \ln(\dot{H}_0/\dot{H})$ is the kinetic coefficient, where $\dot{H} = dH/dt$ and $\dot{H}_0$ is the pre-exponential factor in the Arrhenius relation. The exponents $p$ and $q$ are the fitting parameters describing the dislocation-obstacle interaction profile. Generally speaking, $0 \le p \le 1$ and $1 \le q \le 2$. For the MFIMTs, $p = 1/2$ and $q = 3/2$ are estimated by fitting of Eq. (1) to the magnetic phase diagram of $Ni_{50}Mn_{34.4}In_{15.6}$ which is same as the stress-induced deformation [13,16,20]. Because $\Delta$ and $m$ are positive, the magnetic field hysteresis in this model always increases concomitantly with decreasing temperature. Recently, the magnetic phase diagrams with large hysteresis in several MMSMAs are explained using this model [16–19,21,22].

In this paper, we present the magnetic phase diagrams of the Co-doped Ni-Mn based MMSMAs $Ni_{50-x}Co_xMn_{31.5}Ga_{18.5}$ (x = 9 and 9.7) determined from magnetization measurements in the pulsed high magnetic fields. Temperature dependence of the magnetic field hysteresis shows remarkable reduction below 180 K for $Ni_{41}Co_9Mn_{31.5}Ga_{18.5}$ and below 100 K for $Ni_{40.3}Co_{9.7}Mn_{31.5}Ga_{18.5}$, which cannot be explained by the model described above. The origin of this unusual hysteresis behavior is discussed based on data of magnetizations, microscopy, and X-ray diffraction (XRD) measurements.

Polycrystalline samples of $Ni_{50-x}Co_xMn_{31.5}Ga_{18.5}$ (x = 9 and 9.7) were prepared using the arc melting method. The ingots vacuum-encapsulated in a quartz tube were annealed at 800 °C for 72 hours. Then they were quenched in cold water. High field magnetizations up to 56 T were measured using the nondestructive pulse magnets installed in the Institute for Solid State Physics, the University of Tokyo. To guarantee the isothermal measurements under pulsed fields, the sample was thermally contacted to the large quartz block with silicone grease. Magnetic susceptibility measurements were performed using Quantum Design MPMS with a temperature sweep rate of 2 K/min. Microstructural changes of the sample surface through the MFIMT were observed using the high-speed imaging system installed on the pulse magnet which can generate a magnetic field as strong as 36 T [23]. The XRD patterns were taken using Cu $K_\alpha$ radiation with $\lambda = 1.5418$ Å.

We first took magnetization measurements for $Ni_{41}Co_9Mn_{31.5}Ga_{18.5}$ using pulsed high magnetic fields, to produce the magnetic phase diagram. Fig. 1(a) shows the $M$-$H$ curves and the differential susceptibilities $dM/\mu_0 dH$ of $Ni_{41}Co_9Mn_{31.5}Ga_{18.5}$. The magnetic field up to 56 T is applied three times at 4.2 K after cooling in 0 T. In the first field application, the MFIMTs are observed, respectively, at $H_{M\rightarrow A} = 38.7$ T in the field increasing process and at $H_{A\rightarrow M} = 16.6$ T in the field decreasing process. Here, $H_{M\rightarrow A}$ and $H_{A\rightarrow M}$ are defined as the peaks in $dM/\mu_0 dH$, as presented in Fig. 1(a). Interestingly, $H_{M\rightarrow A}$'s in the second and third field applications are markedly smaller than that of the first field application, although $H_{A\rightarrow M}$ does not change during the three field applications. Fig. 1(b) shows the $M$-$H$ curves measured at various temperatures. Here, the data are taken in order of decreasing temperature from 300 K to 4.2 K. The MFIMTs are observed in all temperature ranges. The $M$-$H$ curves with small hysteresis at low temperatures are obtained as presented in Fig. 1(b), which indicates that the transformation field depends strongly on the thermal and magnetic field history. The magnetic phase diagram and the temperature variation of the hysteresis determined from the $M$-$H$ curves are respectively portrayed in Figs. 1(c) and 1(d). As presented in Fig. 1(d), below room temperature, the hysteresis increases concomitantly with decreasing temperature and shows a peak at 180 K. Subsequently, it decreases concomitantly with decreasing temperature and reaches a constant value below 60 K. This temperature variation of the magnetic field hysteresis of $Ni_{41}Co_9Mn_{31.5}Ga_{18.5}$ is clearly different from that of other Ni-Mn based MMSMAs. It cannot be explained by the model proposed by Umetsu *et al.*, as described above [16–19,21,22].

Next, we applied optical microscopy to the sample surface using the high-speed imaging system in the pulsed magnetic field [23] to elucidate the microstructural origin of the unconventional temperature dependence of the magnetic hysteresis. To lower the transformation field, the extra-Co-doped compound ($Ni_{40.3}Co_{9.7}Mn_{31.5}Ga_{18.5}$) is used. Fig. 2(a) portrays the *M-H* curves of $Ni_{40.3}Co_{9.7}Mn_{31.5}Ga_{18.5}$ measured at 4.2 K. As is the case $Ni_{41}Co_9Mn_{31.5}Ga_{18.5}$, a marked change of the hysteresis is apparent between first and second field applications. Micrographs of the sample surface observed at 4.2 K are shown in Figs. 2(b)-2(d), where the sample surface is polished at a high-temperature A phase to be optically flat. With decreasing temperature, the sample undergoes MT. The surface relief attributable to the martensitic variant appears at low temperatures as presented in Fig. 2(b). Then, the magnetic field of 36 T is applied. Here, the field sweep speed of the pulse magnet used in the optical microscopy is about 4.6 times faster than that used in magnetization measurements. Although the fast field sweep increases the hysteresis of MFIMTs about a few percent [18], the applied magnetic field is sufficiently high to realize the MFIMT. As shown in Figs. 2(c) and (d), it is noteworthy that the overall pattern of the variants does not change during the MFIMTs, although the surface relief disappears partially in the local areas indicated by the red arrows in Fig. 2(c) (i.e. the A phase is partially arrested in the M phase even at 0 T).

The results presented in Figs. 2(b)-2(d) are markedly different from that of NiCoMnIn alloys, which have large magnetic field hysteresis at low temperatures [7,17]. In NiCoMnIn alloys, the surface relief pattern is changed drastically through the MFIMTs, indicating the drastic change of the lattice correspondence variants. This difference of the variants which can be found before and after the MFIMT is explainable by consideration of the total shear strain energy and the Zeeman energy: when the sample undergoes the thermal MT at 0 T with decreasing temperature, the variants are selected to minimize the total shear strain energy in the sample. However, when M phase is realized through the MFIMT with a decreasing magnetic field at low temperature, the variants minimizing the sum of the total shear strain energy and the Zeeman energy are favored. Therefore, if the Zeeman energy is greater than the total shear strain energy, then the variants, of which the easy axis is close to the applied field direction, is selected. Consequently, the pattern of the surface relief differs from that realized at 0 T, such as the case of NiCoMnIn alloys.

When the sample includes the defects such as grain boundaries and microcracks, MT stops at the defects. Therefore, the defects might contribute to the microstructure of the martensitic variants. However, according to our experience, there is no remarkable differences in the grain size and the microcrack density between $Ni_{40.3}Co_{9.7}Mn_{31.5}Ga_{18.5}$ and $Ni_{45}Co_5Mn_{36.7}In_{13.3}$. Thus, the difference of micrographs between the two compounds cannot be explained by the distribution of the defects. Therefore, the results presented in Figs. 2(b)-2(d) might demonstrate that the total shear strain energy is greater than the Zeeman energy in $Ni_{40.3}Co_{9.7}Mn_{31.5}Ga_{18.5}$. The difference of the shear strain energy between NiCoMnGa and NiCoMnIn alloys might be explained by considering the structural compatibility between A and M phases in the habit plane. If the lattices of A and M phases match perfectly in the habit plane, then the elastic energy in the habit plane is zero. If not, then the finite elastic energy is stored: it functions as an obstacle to the habit plane motion. According to theoretical calculations, the 2M structure of M phase (NiCoMnGa) has greater structural incompatibility in the habit plane than 10M and 14M structures (NiCoMnIn) [24]. Therefore, in NiCoMnGa alloys, greater Zeeman energy is needed for the habit plane motion than that of NiCoMnIn alloys.

As described above, modulation in the M phase plays an important role in the MT of Ni-Mn based MMSMAs. For that reason, we specifically examine the modulation of M phase. It has been reported theoretically that the energetically reasonable paths of thermal MT in stoichiometric $Ni_{50}Mn_{25}Ga_{25}$ are the following: A → 6M (pre-M phase) → 10M or 14M → 2M, which is also confirmed experimentally [25–27]. Recently, Konoplyuk *et al.* reported that non-stoichiometric $Ni_{51.9}Mn_{27}Ga_{21.1}$ also undergoes the thermal MT as follows: A → 14M → 2M with decreasing temperature [28]. They also report that, because of the difference of the transformation path in the free energy profile, the thermal hysteresis depends strongly on the thermal history. However, such inter-martensitic structural changes are not observed in $Ni_{41}Co_9Mn_{31.5}Ga_{18.5}$ (Supplemental Information).

Magnetic susceptibility measurements are taken to obtain more detailed information about the thermal and the magnetic field history dependence of the hysteresis. As shown in Figs. 3(a) and 3(b), $Ni_{40.3}Co_{9.7}Mn_{31.5}Ga_{18.5}$ produces similar *M-H* curves and magnetic phase diagrams to $Ni_{41}Co_9Mn_{31.5}Ga_{18.5}$, in which the magnetic field hysteresis decreases concomitantly with decreasing temperature below 100 K. It is noteworthy that $H_{M \to A}$ and $H_{A \to M}$ in $Ni_{40.3}Co_{9.7}Mn_{31.5}Ga_{18.5}$ are lower than that of $Ni_{41}Co_9Mn_{31.5}Ga_{18.5}$. Therefore, $Ni_{40.3}Co_{9.7}Mn_{31.5}Ga_{18.5}$ remains in the A phase at low temperatures through the field cooling above 7 T, which can be reached using MPMS. Therefore, we used $Ni_{40.3}Co_{9.7}Mn_{31.5}Ga_{18.5}$ for magnetic susceptibility measurements.

Fig. 4(a) shows the $\chi$–$T$ curve for $Ni_{40.3}Co_{9.7}Mn_{31.5}Ga_{18.5}$ measured at 0.01 T. The solid and open circles respectively represent results measured during cooling and heating processes, where the measurements are performed as follows: 350 K → 5 K → 350 K. In both processes, the steep susceptibility changes are observed at around 250 K and 300 K. They are attributed respectively to the Curie temperature of M phase ($T_C^M$) and the MT temperatures [$T_{M \to A}$ (heating process), and $T_{A \to M}$ (cooling process)]. $T_C^M = 254.2$ K is estimated by the dip of $d\chi/dT$, as portrayed in Fig. 4(b). Similarly, $T_{M \to A} = 304.3$ K and $T_{A \to M} = 273.9$ K are estimated (i.e. the hysteresis of the thermal MT is $T_{Hys} = 30.4$ K). Open triangles in Fig. 4(a) show the $\chi$–$T$ curve measured by increasing the temperature from 5 K to 350 K after the field cooling at 7 T. Here, the magnetic field of 7 T is applied at 350 K;

then the sample is cooled to 5 K at 7 T. Subsequently, the magnetic field is removed. Thereby, the sample undergoes the MFIMT from A to M phase before the measurement. It is noteworthy that the estimated $T_{M \rightarrow A} = 298.4$ K (i.e. $T_{Hys} = 24.5$ K) is remarkably lower than when the sample undergoes thermal MT at 0.01 T. This result implies that the microstructure of M phase realized through the MFIMT differs from that realized through the thermal MT.

To understand the thermal and magnetic field history dependence of the hysteresis of $Ni_{50-x}Co_xMn_{31.5}Ga_{18.5}$ ($x = 9$ and $9.7$), herein we propose a simple two-variant model illustrated schematically in Figs. 4(c-1)-4(c-6). Fig. 4(c-2) portrays the 2M structure of M phase which is realized in these compounds at low temperatures through the zero field cooling. The arrows in each variant indicate the magnetic moments ($\boldsymbol{M}$), which are aligned with the easy axes of the two twin variants. In the first field application, the MFIMT from M to A phase with large $H_{M \rightarrow A}$ is realized as shown as the red solid curve in Fig. 4(c-1). In this case, the variants are removed through the MFIMT as follows: Figs. 4(c-2) $\rightarrow$ 4(c-3) $\rightarrow$ 4(c-4). Subsequently, the sample undergoes MFIMT from A to M phase in the field decreasing process. Here, if the magnetic field is applied parallel to the easy axis of one variant, then the Zeeman energy difference between the two variants is described as $M\mu_0 H \cos\varphi$, where $\varphi$ represents the angle between easy axes of the two twin variants. Because the NiMnGa alloys have strong magnetic anisotropy [29–31], the Zeeman energy difference assists in stabilizing one variant. In this case, the modulated structures such as 10M and 14M might be more stable than the 2M structure. This process can be represented as follows: Figs. 4(c-4) $\rightarrow$ 4(c-5) $\rightarrow$ 4(c-6). If the modulated structure created through MFIMT remains at 0 T, then in the second field application [blue dashed curve in Fig. 4(c-1)], the sample undergoes MFIMT from modulated M to A phase [Figs. 4(c-6) $\rightarrow$ 4(c-5) $\rightarrow$ 4(c-4)]. According to the theoretical calculations [26,27], in this case, the sample undergoes MFIMT through a different transformation path with a smaller energy barrier (i.e. lower $H_{M \rightarrow A}$). Similarly, when the sample is heated and undergoes the thermal MT, the MT is realized from modulated M to A phase. Therefore, the $T_{M \to A}$ can be decreased. If the modulated structures are realized all over the sample, then the microtwin variants change drastically; they must be detected using microscopy as a change of the surface relief. However, as shown in Figs. 2(b)-2(d), no changes are observed within the resolution of the microscope. This lack of change might imply the partial realization of modulated structures. Because the modulated structure is the metastable state, relaxation to the 2M structure can occur, when the measurement temperature is sufficiently high to overcome the energy barrier between the two structures. Consequently, the volume fraction of the modulated structure depends strongly on the measurement temperature. The temperature dependence of the hysteresis with a peak at 180 K shown in Fig. 1(d) might imply that the volume fraction of the modulated structures increases with decreasing temperature below 180 K.

Summarizing, magnetic phase diagrams of $Ni_{50-x}Co_xMn_{31.5}Ga_{18.5}$ (x = 9 and 9.7) are found. For both compounds, unconventional magnetic field hysteresis, which decreases with decreasing temperature after showing a peak, is observed. Furthermore, the thermal MTs are investigated through magnetic susceptibility measurements for $Ni_{40.3}Co_{9.7}Mn_{31.5}Ga_{18.5}$. Through field cooling at 7 T, the hysteresis of the thermal MT decreases about 20 %. The thermal and magnetic field hysteresis are understood comprehensively by the model considering the magnetic field induced modulation of M phase. Microscopic experiments such as X-ray diffraction in the magnetic fields are expected to clarify the microstructure of the field-induced M phase in $Ni_{50-x}Co_xMn_{31.5}Ga_{18.5}$ (x = 9 and 9.7).

**Acknowledgment**

This work was partly supported by the Ministry of Education, Culture, Sports, Science, and Technology, Japan, through a Grant-in-Aid for Early Career Scientist (Grant No. 18K13979). Some of this work was performed through the joint research with the Institute for Solid State Physics, The University of Tokyo.

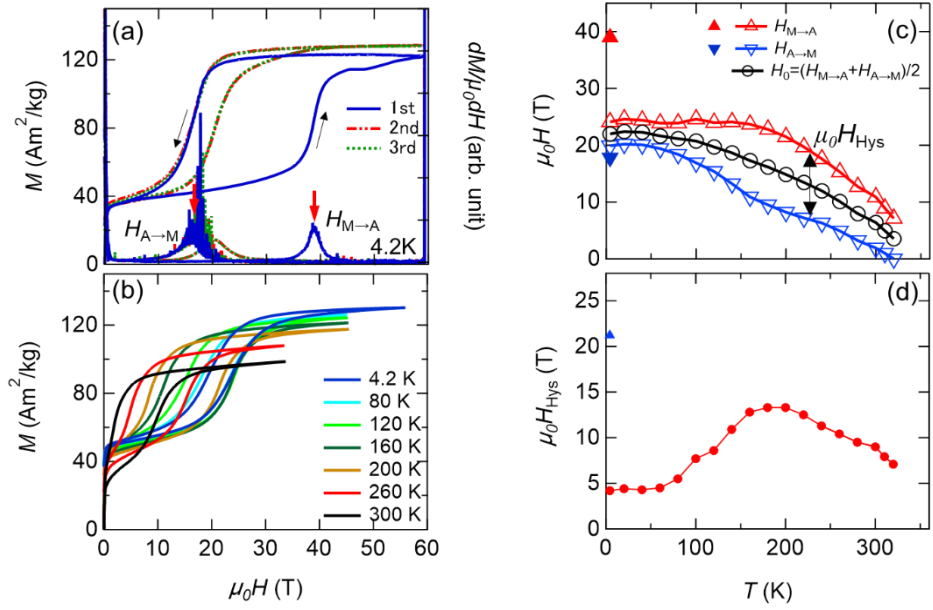

Fig. 1 (a) Magnetic field dependence of magnetization and differential susceptibility $dM/\mu_0 dH$ of Ni$_{41}$Co$_9$Mn$_{31.5}$Ga$_{18.5}$ measured at 4.2 K. Red-downward arrows represent, respectively, the field induced martensitic transformation fields ($H_{M\rightarrow A}$: field increasing process and $H_{A\rightarrow M}$: field decreasing process). (b) $M$-$H$ curves measured at the various temperatures. (c) Magnetic phase diagram determined by magnetization measurements. The solid triangles respectively denote $H_{M\rightarrow A}$ and $H_{A\rightarrow M}$ in the $M$-$H$ curve of the first field application. (d) Temperature dependence of the magnetic hysteresis. The solid triangle is the hysteresis in the first field application.

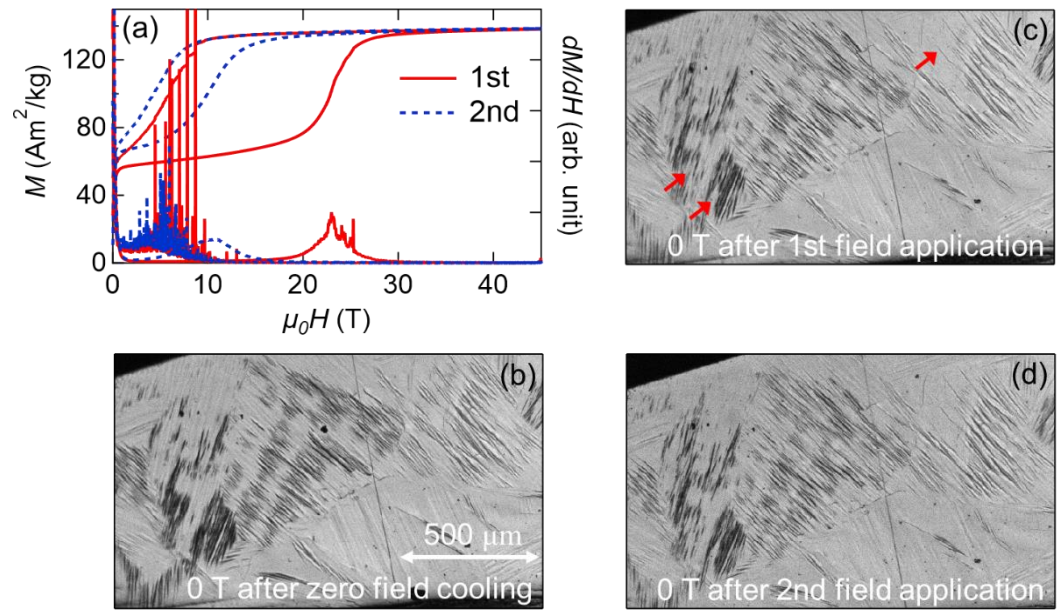

Fig. 2 (a) *M-H* curves of Ni$_{40.3}$Co$_{9.7}$Mn$_{31.5}$Ga$_{18.5}$ measured at 4.2 K. (b)-(d) Micrographs of martensitic phase of Ni$_{40.3}$Co$_{9.7}$Mn$_{31.5}$Ga$_{18.5}$ observed at 8.4 K [(b) after zero field cooling, (c) after first field application, (d) after second field application].

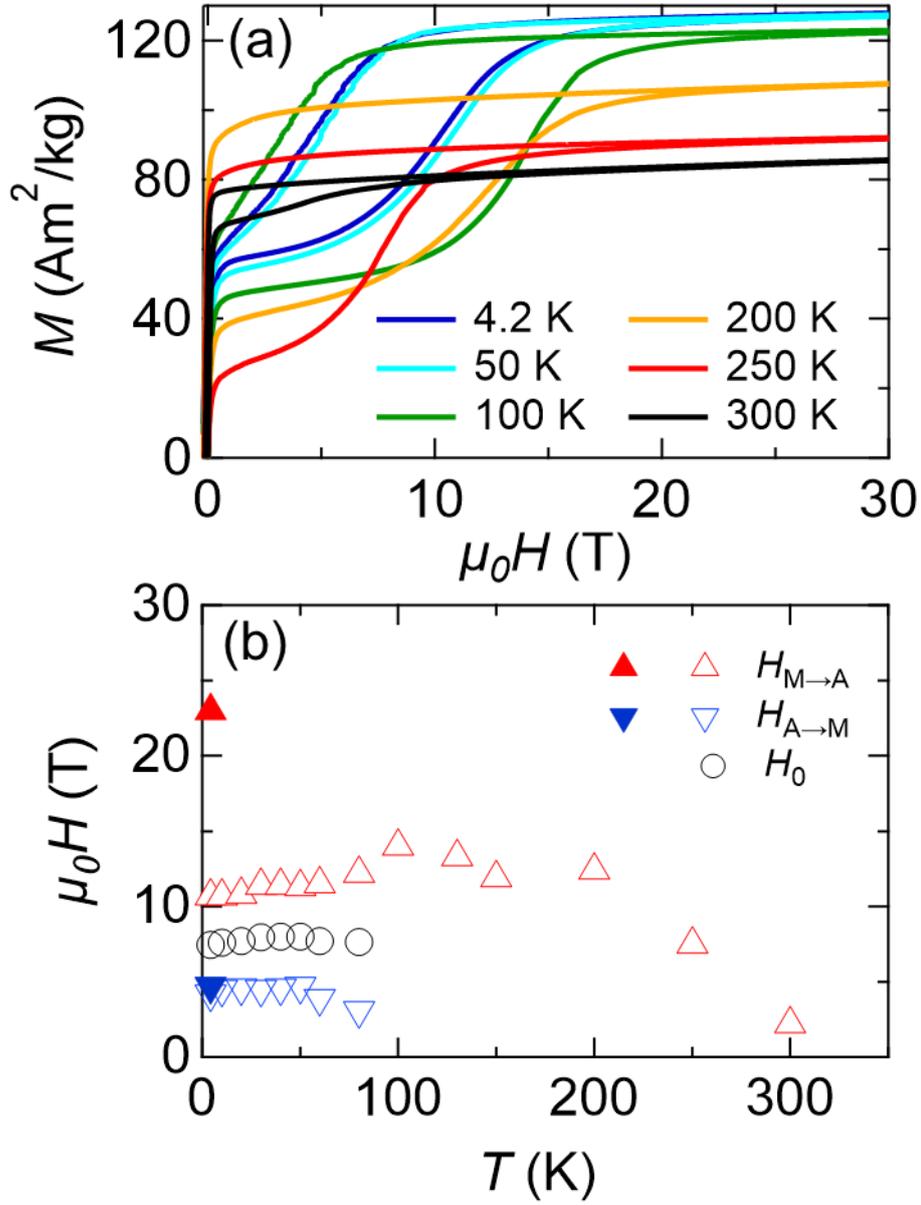

Fig. 3 (a) *M-H* curves of Ni$_{40.3}$Co$_{9.7}$Mn$_{31.5}$Ga$_{18.5}$ measured at various temperatures. (b) Magnetic phase diagram determined by the *M-H* curves, where $H_{M \to A}$, $H_{A \to M}$ and $H_0$ are defined in Fig. 1(a). The solid triangles respectively denote $H_{M \to A}$ and $H_{A \to M}$ in the *M-H* curve of the first field application shown in Fig. 2(a).

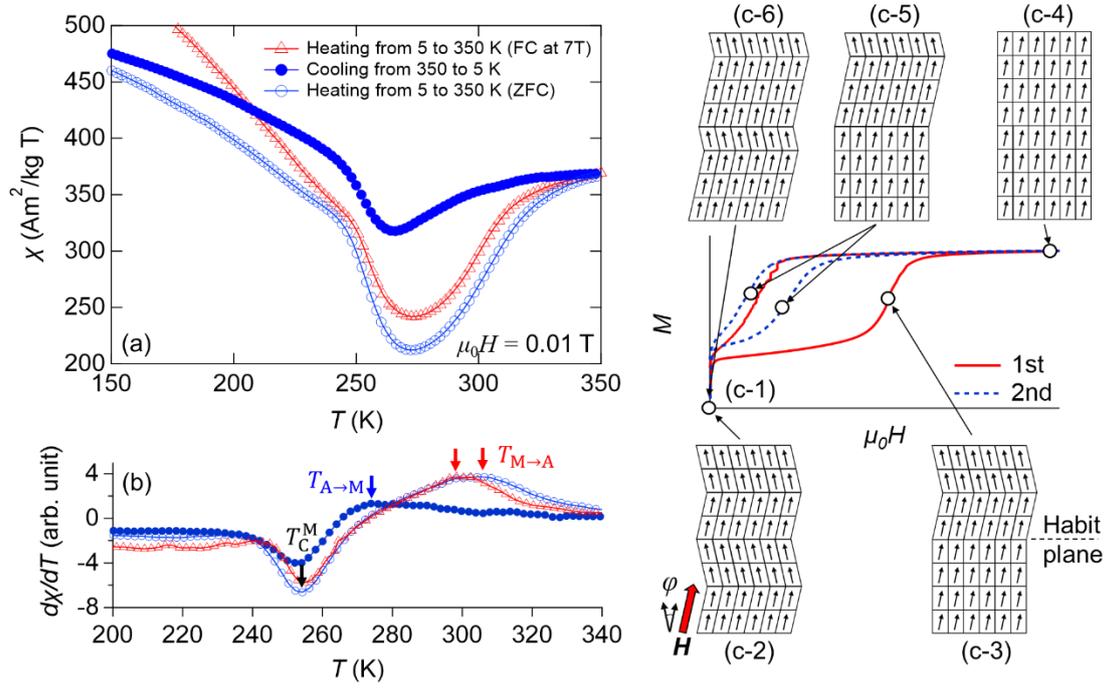

Fig. 4 (a) Temperature dependence of magnetic susceptibility of $Ni_{40.3}Co_{9.7}Mn_{31.5}Ga_{18.5}$ measured at 0.01 T. Open and solid symbols respectively denote heating and cooling processes. Open triangles are measured after the field cooling at 7 T. (b) Temperature differential of magnetic susceptibility. Transformation temperatures and the Curie temperature of M phase are determined as the peaks. (c-1) $M$-$H$ curves of $Ni_{40.3}Co_{9.7}Mn_{31.5}Ga_{18.5}$ measured at 4.2 K. (c-2)-(c-6) Schematic illustration of the change in the microstructure through the MFIMTs.

# Supplemental information for X-ray diffraction for $Ni_{41}Co_9Mn_{31.5}Ga_{18.5}$

T. Kihara[1]*, X. Xu[2], A. Miyake[3], Y. Kinoshita[3] M. Tokunaga[3], Y. Adachi[4], and T. Kanomata[5]

[1] Institute for Materials Research, Tohoku University, Sendai, Miyagi, Japan

[2] Department of Materials Science, Tohoku University, Sendai, Miyagi, Japan

[3] The Institute for Solid State Physics, The University of Tokyo, Kashiwa, Chiba, Japan

[4] Graduate School of Science and Engineering, Yamagata University, Yonezawa, Yamagata, Japan

[5] Research Institute for Engineering and Technology, Tohoku Gakuin University, Tagajo, Miyagi, Japan

* Address all correspondence to: t_kihara@imr.tohoku.ac.jp

This supplemental information provides X-ray diffraction (XRD) results of $Ni_{41}Co_9Mn_{31.5}In_{18.5}$. The XRD patterns were obtained using Cu $K_\alpha$ radiation. Here, we used a plate shaped polycrystalline sample, which is made from the same ingot as the sample used for the magnetization measurements. The X-ray beam was applied to the polished sample surface. The incident beam is shaped by the 2 mm slit before hitting the sample.

Figure S-1 shows the XRD patterns measured at various temperatures. All peaks observed in the angular range of $2\theta$ from 30° to 115° are well assigned to the reflections from 2M martensitic phase. Here, the peaks indicated by arrows are attributable to the niobium gasket, which is located in the sample chamber on the X-ray diffractometer. No other peaks are attributable to modulated martensitic phases such as 6M, 10M, and 14M martensite. In addition, as shown in Fig. S-2, the temperature dependence of intensity of the peaks attributable to the 2M martensite shows no anomaly, which indicates that the absence of such modulated martensite in this temperature range.

According to the specific heat measurements taken using a differential scanning calorimeter, the martensitic transformation temperature ($T_{M \to A}$) in the temperature increasing process is determined as $T_{M \to A}$ = 364 K (data are not presented). Furthermore, the XRDs are measured up to 333 K, which is indeed close to $T_{M \to A}$. Therefore, we concluded that $Ni_{41}Co_9Mn_{31.5}In_{18.5}$ undergoes the martensitic transformation directly from cubic austenitic phase to 2M martensitic phase. The ground state is 2M martensite.

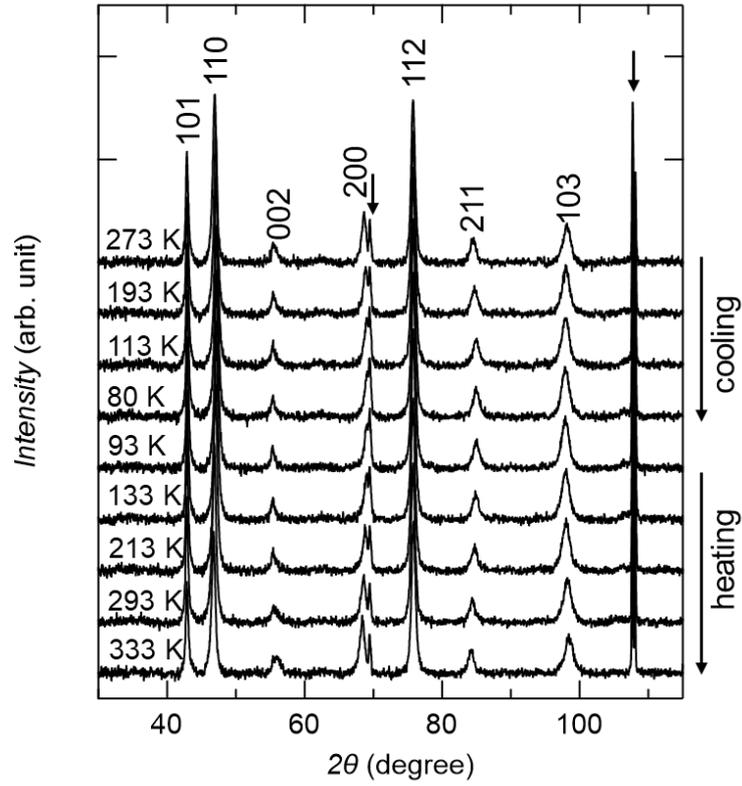

Fig. S-1 Temperature variation of XRPD pattern of $Ni_{41}Co_9Mn_{31.5}In_{18.5}$ in cooling and heating processes.

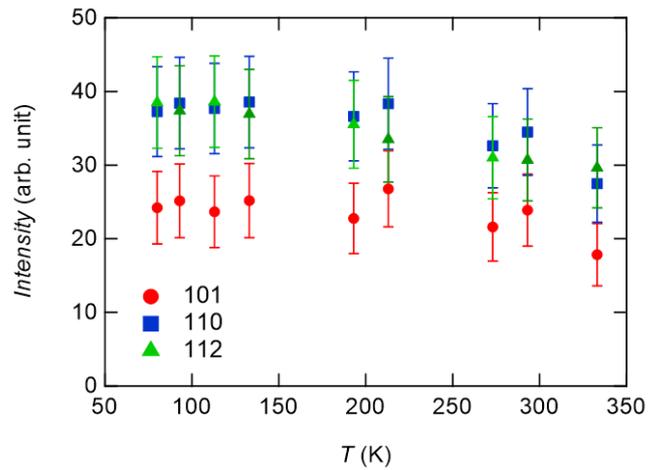

Fig. S-2 Temperature dependence of the intensity of (1 0 1), (1 1 0), and (1 1 2) peaks.